\documentclass[12pt]{iopart}

\expandafter\let\csname equation*\endcsname=\relax
\expandafter\let\csname endequation*\endcsname=\relax
\usepackage{amsmath}
\usepackage{graphicx} 
\usepackage{amsmath,amsfonts,
	amsgen,amsbsy,amsopn,amstext,amssymb
}
\usepackage{caption}
\usepackage{epstopdf}
\usepackage{subcaption}
\usepackage[
colorlinks=true,
linkcolor=blue,
citecolor=blue,
filecolor=blue,
urlcolor=blue]{hyperref}
\usepackage{cite}

\begin{document}

\title[Limits of flexural wave absorption by open lossy resonators]{Limits of flexural wave absorption by open lossy resonators: reflection and transmission problems}

\author{J. Leng$^1$, F. Gautier$^1$, A. Pelat$^1$, R. Pic\'o$^2$, J.-P. Groby$^1$, \\V. Romero-Garc\'ia$^1$}

\address{$^1$Laboratoire d'Acoustique de l'Universit\'e du Mans, LAUM - UMR 6613 CNRS, Le Mans Universit\'e, Avenue Olivier Messiaen, 72085 LE MANS CEDEX 9, France\newline
$^2$Instituto para la Gesti\'on Integral de zonas Costeras (IGIC), Universitat Polit\`ecnica de Val\`encia, Paranimf 1, 46730, Gandia, Spain}
\ead{vicente.romero@univ-lemans.fr}

\begin{abstract}
The limits of flexural wave absorption by open lossy resonators are analytically and numerically reported in this work for both the reflection and transmission problems. An experimental validation for the reflection problem is presented. The reflection and transmission of flexural waves in 1D resonant thin beams are analyzed by means of the transfer matrix method. The hypotheses, on which the analytical model relies, are validated by experimental results. The open lossy resonator, consisting of a finite length beam thinner than the main beam, presents both energy leakage due to the aperture of the resonators to the main beam and inherent losses due to the viscoelastic damping. Wave absorption is found to be limited by the balance between the energy leakage and the inherent losses of the open lossy resonator. The perfect compensation of these two elements is known as the critical coupling condition and can be easily tuned by the geometry of the resonator. On the one hand, the scattering in the reflection problem is represented by the reflection coefficient. A single symmetry of the resonance is used to obtain the critical coupling condition. Therefore the perfect absorption can be obtained in this case. On the other hand, the transmission problem is represented by two eigenvalues of the scattering matrix, representing the symmetric and anti-symmetric parts of the full scattering problem. In the geometry analyzed in this work, only one kind of symmetry can be critically coupled, and therefore, the maximal absorption in the transmission problem is limited to 0.5. The results shown in this work pave the way to the design of resonators for efficient flexural wave absorption.
\end{abstract}

\maketitle

\section{Introduction} 

Recent studies in audible acoustics have focused on wave absorption at low frequencies by means of subwavelength locally resonant materials. In particular, the design of broadband subwavelength perfect absorbers, whose dimensions are much smaller than the wavelength of the frequency to be attenuated, has recently been proposed \cite{romero2016use,merkel2015control,groby2016use,jimenez2016ultra,jimenez2017rainbow}. Such devices can totally absorb the energy of an incident wave and require solving the twofold but often contradictory problem: ($i$) increasing the density of states at low frequencies and ($ii$) matching the impedance with the background medium. On the one hand, the use of local resonators is a successful approach for increasing the density of states at low frequencies with reduced dimensions, as it has been shown in the field of metamaterials \cite{fang2006ultrasonic,liu2000locally,skelton18,Wei14,Duan15, theocharis2014, Colombi17,Pal17}. On the other hand, the local resonators of such metamaterials are open and lossy ones, implying energy leakage and inherent losses. In these systems the impedance matching can be controlled by the ratio between the inherent losses of the resonator and the leakage of energy \cite{bliokh2008colloquium}. Particularly, the perfect compensation of the leakage by the losses is known as the critical coupling condition \cite{yariv2000universal} and has been widely used to design perfect absorbers in different fields of physics\cite{xu2000scattering,cai2000observation} other than acoustics.

The critical coupling condition is also relevant for applications in vibrations owing to the increasing need for damping materials at low frequencies in several branches of industry \cite{Duan15}. Current passive solutions in this field are mainly based on the use of viscoelastic coatings \cite{teng2001analysis}. Another solution yields in  the tuned vibration absorber (TVA) \cite{den1985mechanical, Brennan99, El-Khatib05} that is used to control flexural waves in beams. The tuning of the resonance frequency of an undamped TVA has been analyzed \cite{Brennan99}, showing that complete suppression of the flexural wave transmission can be achieved. In most cases, TVA have been used to minimize the transmission of a propagating wave \cite{Brennan99}, resulting in practice in heavy treatments at low frequencies. Less attention has been paid to the case of maximizing the absorption in order to reduce simultaneously both the reflected and transmitted waves. 

The purpose of this work is to study the problem of perfect absorption of flexural waves in 1D elastic beams with local resonators by using the critical coupling condition. Particularly, the absorption of energy is analyzed through the balance between the energy leakage and the inherent losses in the resonators for the two scattering problems: the reflection and the transmission of flexural waves. The presented problem is related to the control of flexural waves in a beam using a passive TVA but with a physical insight that allows the interpretation of the limits of the flexural wave absorption based on both the critical coupling conditions and the symmetry of the excited resonances in the resonator. The analyzed systems are composed of a main beam and an open resonator simply consisting in a reduction of the thickness of the main beam. A thin viscoelastic coating is attached to it, leading to a composite material whose loss may be tuned. This composite material is modeled with the Ross-Kerwin-Ungar (RKU) method \cite{rossdamping} and is embedded in the main beam. By tuning the losses, it is possible to analyze the different limits in both scattering problems. In practice, this type of resonator results in simpler geometries than that of the TVA which consists of complicated combinations of mass spring systems simulating a point translational impedance.

The composite is studied by means of an analytical model based on the transfer matrix method. The analytical results, in accordance with the the experimental results, show the limits of the maximal values for the flexural wave absorption and their physical interpretations in both the reflection and transmission problems. The interpretations are based on the eigenvalues of the $S$-matrix for the propagating waves, represented in the complex frequency plane \cite{romero2016use}. An experimental prototype is designed and measured for the reflection problem. The experimental results prove the perfect absorption of flexural waves and validate the analytical predictions.

The work is organized as follows. In section \ref{sec:theo}, the theoretical model used to analyze the 1D scattering problems of flexural wave is presented. The physical analysis of the absorption coefficient in the complex frequency plane are presented in section \ref{sec}. This analysis is based on an analytical model and the concept of critical coupling to obtain a perfect absorption of flexural waves. The experimental set-up used to validate the model for the reflection problem is then presented in section \ref{sec:exp1} as well as the experimental methodology and results. Finally, section \ref{sec:con} summarizes the main results and gives the concluding remarks.

\section{Theoretical models}
\label{sec:theo}

\begin{figure}
\includegraphics[width=16cm]{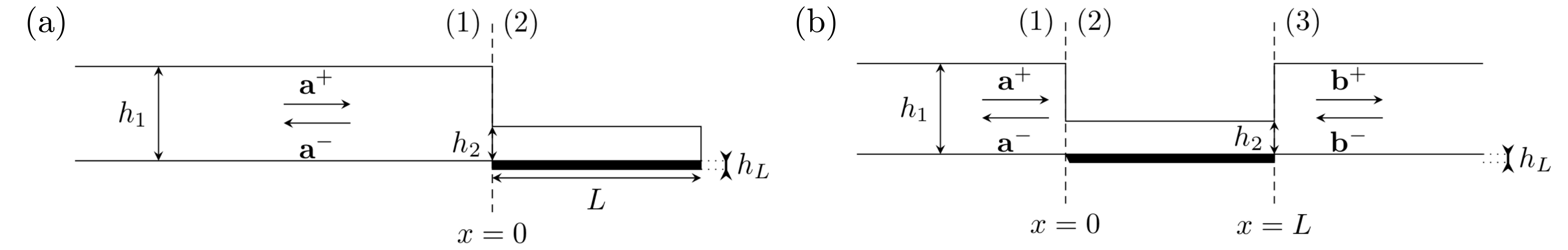}
\caption{Diagrams of the 1D configurations analyzed for the reflection and transmission problems for flexural waves. (a) Configuration for the reflection problem. (b) Configuration for the transmission problem. }
\label{fig:fig1}
\end{figure}

This section describes the theoretical model used to study the absorption of flexural waves by open lossy resonators in 1D systems, following the approach of Mace \cite{mace1984wave}. The governing equations used in the model are first introduced. Two scattering problems are then presented. The first one is the reflection problem where the absorption by a resonator made of a thinner composite beam located at the termination of a semi-infinite beam is studied (figure~\ref{fig:fig1}a). The second one is the transmission problem where the absorption of the same resonator located between two semi-infinite beams is considered (figure~\ref{fig:fig1}b). The analytical results shown for the two problems have been tested by numerical simulations,  but not shown in the article for clarity of the figures, later on the analytical results are validated experimentally in the section \ref{sec:exp1}.

\subsection{Flexural wave propagation in uniform beams}
Consider a thin uniform beam whose neutral axis is denoted by the $x$-axis. Assuming Euler-Bernoulli conditions, the flexural displacement $w(x,t)$ satisfies\cite{leissa1969vibration}:
\begin{equation}
\label{eq1}
D\frac{\partial^4 w}{\partial x^4}+ m \frac{\partial^2 w}{\partial t^2}= 0,
\end{equation}
where $D=EI$ is the flexural rigidity, $E$ the Young modulus, $I$ the second moment of area and $m$ the linear mass. Assuming time harmonic solution of the form $e^{i \omega t}$, where $\omega$ is the angular frequency, the solution of Eq.(\ref{eq1}) can be written in the frequency domain as the sum of four flexural waves:
\begin{equation}
w(x)= a^+ e^{-ikx}+a_N^+ e^{-kx}+a^- e^{ikx}+a_N^- e^{kx}.
\end{equation}
The complex amplitudes of the propagative and evanescent waves are $a$ and $a_N$ respectively, and the signs $^+$ and $^-$ denote the outgoing and ingoing waves respectively. The evanescent component is a near field component, the amplitude of which decreases exponentially with distance. The flexural wavenumber $k$ is given by $k^4=\displaystyle{\frac{m \omega^2}{D}}$, 
which is real and positive in the lossless case and complex when damping is accounted for. The wave amplitude is expressed in the vector form by convenience:
\begin{equation}
\mathbf{a}^+=\left[\begin{matrix}
a^+\\
a_N^+
\end{matrix}\right],\;\;\;\;\;
\mathbf{a}^-=\left[\begin{matrix}
a^-\\
a_N^-
\end{matrix}\right].
\end{equation}
The relation between wave amplitudes along a beam with a constant thickness are then described by
\begin{equation}
\mathbf{a}^+(x_0+x)= \mathbf{f} \mathbf{a}^+(\mathbf{x_0})\; \;  \textrm{and}\; \;  \mathbf{a}^-(x_0+x)= \mathbf{f}^{-1} \mathbf{a}^-(\mathbf{x_0}),
\end{equation}
where the diagonal transfer matrix $\mathbf{f}$ is given by
 \begin{equation}
 \mathbf{f}=\left[\begin{matrix}
 e^{-ikx} & 0\\
 0 &  e^{-kx}
 \end{matrix}\right].
 \end{equation}

\subsection{Reflection coefficient in a pure reflection problem}

Consider an incident plane wave in the configuration described by the figure~\ref{fig:fig1}a, where the system is terminated by a free termination at one end. The displacement $\mathbf{w}$ at any point for $x<0$ reads as
\begin{equation}
\mathbf{w}(x<0)= \mathbf{a^+}+\mathbf{a^-} = \mathbf{a^+}+\mathbf{R_r}\cdot\mathbf{a^+},
\end{equation}
where $\mathbf{R_r}$ denotes the reflection matrix of the resonant termination of the beam at $x=0$. The incident wave is transmitted into the resonant termination and reflected at its end, therefore the matrix $\mathbf{R}_r$ can be evaluated, using the displacement continuity at the interface and at the boundaries as \cite{mace1984wave}: 
\begin{equation}
\mathbf{R_r}= \mathbf{a^-}\mathbf{a^+}^{-1} = \mathbf{r_{12}}+\mathbf{t_{12}}\left((\mathbf{f} \mathbf{r_f} \mathbf{f})^{-1} -\mathbf{r_{21}}\right)^{-1}\mathbf{t_{21}},
\end{equation}
where $\mathbf{r_{ij}}$ and $\mathbf{t_{ij}}$ represent the reflection and transmission matrices from section $(i)$ to section $(j)$ of the beam (see figure~\ref{fig:fig1}a). Considering continuity and equilibrium respectively at the section change, these matrices are given by
\begin{equation}
 \mathbf{t_{ij}}= \frac{4}{\Delta	} \left[\begin{matrix}
 (1+\beta)(1+\gamma) & (-1+1i\beta)(1-\gamma)\\
 (-1-i\beta)(1-\gamma) & (1+\beta)(1+\gamma)
 \end{matrix}\right],
\end{equation}
\begin{equation}
 \mathbf{r_{ij}}= \frac{2}{\Delta	} \left[\begin{matrix}
 -2(\beta^2-1)\gamma-1i\beta(1-\gamma)^2 & (1+i)\beta(1-\gamma^2)\\
 (1-i)\beta(1-\gamma^2) & -2(\beta^2-1)\gamma+i\beta(1-\gamma)^2
 \end{matrix}\right],
\end{equation}
where $\displaystyle{\beta=\frac{k_j}{k_i}}$ and $\displaystyle{\gamma=\frac{D_jk_j^2}{D_ik_i^2}}$ correspond to the ratios of wavenumbers and bending wave impedances, and $\Delta= (1+\beta)^2(1+\gamma)^2-(1+\beta^2)(1-\gamma)^2$. The reflection matrix $\mathbf{r_{f}}$ of the free termination is given by
\begin{equation}
\mathbf{r_f}=\left[\begin{array}{cc} -\imath & (1+\imath)\\(1-\imath) & \imath \end{array}\right].
\end{equation}
 $\mathbf{R_r}$ is thus a $2\times 2$ matrix where the diagonal components correspond to the reflection coefficients of the propagative and evanescent waves respectively. The study focuses on the reflection of waves in the far-field ($x\rightarrow -\infty$), i.e., on the propagative waves that carry the energy. The first term of the reflection matrix $\mathbf{R_r}(1,1)\equiv R_r$ is therefore only considered since $\mathbf{R_r}(1,2),\mathbf{R_r}(2,1),\mathbf{R_r}(2,2)\rightarrow 0$ when $x\rightarrow -\infty$. The absorption coefficient $\alpha_R$ of propagating waves in the reflection problem can then be written as:
 \begin{equation}
 \alpha_{r} = 1- \vert R_r \vert^2,\;\;\;\;\;\;x\rightarrow -\infty.
 \end{equation}
In the lossless case, i.e. without dissipation, $R_r$ is simply equal to 1 for any purely real frequency as the energy conservation is fulfilled.

\subsection{Reflection and transmission coefficients in a 1D symmetric and reciprocal transmission problem}
The transmission problem of the structure shown in figure~\ref{fig:fig1}b is described in this section, considering $\mathbf{b^-}=0$. Due to the symmetry of the resonator and assuming propagation in the linear regime, the problem is considered as symmetric and reciprocal. The reflection and transmission matrices $\mathbf{R_t}$ and $\mathbf{T_t}$ at $x=0$ and $x=L$ are used to define the displacements on each side of the resonator such as:
\begin{equation}
\mathbf{w}{(x<0)}= \mathbf{a^+}+\mathbf{a^-} = \mathbf{a^+}+\mathbf{R_t}\cdot\mathbf{a^+},
\end{equation}
\begin{equation}
\mathbf{w}{(x>L)}= \mathbf{b^+}= \mathbf{T_t}\cdot\mathbf{a^+}.
\end{equation}
 Using the displacement continuity at $x=0$ and $x=L$ in a similar way as in the previous section, $\mathbf{R_t}$ and $\mathbf{T_t}$ are written as
\begin{equation}
\mathbf{R_t}= \mathbf{r_{12}}+\mathbf{t_{12}}\left((\mathbf{f} \mathbf{r_{23}} \mathbf{f})^{-1} -\mathbf{r_{21}}\right)^{-1}\mathbf{t_{21}},
\end{equation}
\begin{equation}
\mathbf{T_t}= \mathbf{a^-}\mathbf{a^+}^{-1} = \mathbf{t_{23}} (\mathbf{I} - (\mathbf{f} \mathbf{r_{21} \mathbf{f})} \mathbf{r_{23}})^{-1}\mathbf{f} \mathbf{t_{12}}.
\end{equation}
Therefore the absorption coefficient of the transmission problem is defined as
 \begin{equation}
 \alpha_{t} = 1- \vert T_t \vert^2 - \vert R_t \vert^2,\;\;\;\;\;x\rightarrow \pm\infty
 \end{equation}
where $R_t=\mathbf{R_t}(1,1)$ when $\;x\rightarrow -\infty$ and $T_t=\mathbf{T_t}(1,1)$ when $\;x\rightarrow +\infty$.

\subsection{Viscoelastic losses in the resonator: the RKU model}

The inherent losses of the resonator are introduced by a thin absorbing layer of thickness $h_l$ as shown in figures~\ref{fig:fig1}a-\ref{fig:fig1}b and are considered frequency independent. The complex Young Modulus of the absorbing layer is $E_{l}(1 + i\eta_l)$, where $\eta_l$ is its loss factor. Using the RKU model\cite{rossdamping}, this region is modeled as a single composite layer with a given effective bending stiffness $D_c$ written as:
\begin{equation}
D_c=E_2I_2\left[(1+j\eta_2)+e_ch_c^3(1+j\eta_l)+\frac{3+(1+h_c)^2e_ch_c[1-\eta_2\eta_l+j(\eta_2+\eta_l)]}{1+e_ch_c(1+j\eta_l)}\right] ,
\end{equation}
where the indices $2$ and $l$ stand for the parameters of the thin beam and of the absorbing layer respectively, $e_c=E_l/E_2$ and $h_c=h_l/h_2$. The wave number $k_c$ of the composite material can then be written as $k_c^4=\displaystyle{\frac{\rho_c h\omega^2}{D_c}}$, where $h=h_l+h_2$ and $\rho_c h=\rho_2h_2+\rho_lh_l$.

\section{Limits of absorption for the reflection and transmission problems}
\label{sec}

This section describes the limits of absorption for flexural waves in the reflection and transmission problem by using open, lossy and symmetric resonators. It provides tools to design absorbers with a maximal absorption in both problems. For this purpose, the eigenvalues of the scattering matrix of the propagative waves are represented in the complex frequency plane as in Ref. \cite{romero2016use}. The material and geometric parameters used in the following sections are described in table \ref{table1}.

\begin{table}
\lineup
\caption{\label{table1}Geometric and material parameters of the studied systems. The value of $\eta_l$ depends on the experimental set-up used, see main text for the used values retrieved from experiments.}
\begin{indented}
\item[]\begin{tabular}{@{}llll}
\br
&Geometric parameters & Material parameters\\
\mr
Main beam & $h_1=5$~mm & $\rho=2811$~kg.m$^{-3}$ \\
				 & $b = 2$~cm & $E=71.4$~GPa \\
 				 &					& $\eta=0$\\
				 &					& $\nu=0.3$\\
				 & & \\
Resonator beam & $h_2=0.217$~mm & $\rho_2=2811$~kg.m$^{-3}$ \\
						 &$b_2=2$~cm& $E_2=71.4$~GPa \\
    		 				 & $L=1.6$~cm					& $\eta_2=0$\\
						 &  & $\nu_2=0.3$ \\
						 & & \\
Coating layer & $h_l=1.5$~mm &$E_l=6.86\times 10^{-3}$~GPa \\
					& $b_l=2$~cm & $\rho_l=93.3$~kg.m$^{-3}$\\
					& $L_l=1.6$~cm & $\eta_l$ \\
					& 	 & $\nu_l=0.3$ \\
\br
\end{tabular}
\end{indented}
\end{table}

\subsection{Properties of the S-matrix}
\label{Property_S_matrix}

Consider a two-port, 1D, symmetric and reciprocal scattering process for the systems described in figure~\ref{fig:fig1}b in the case where $x\rightarrow\pm\infty$. The relation between the amplitudes $a^+$ and $b^-$ of the incoming waves and $a^-$ and $b^+$ of the outgoing waves when $x\rightarrow\pm\infty$ is given by
\begin{equation}
\left( \begin{matrix}  a^- \\ b^+ \end{matrix}\right)= \mathbf{S}(\omega) \left( \begin{matrix}  b^- \\ a^+ \end{matrix}\right)=\left[ \begin{matrix}
T_t & R_t \\ R_t & T_t
\end{matrix} \right] \left( \begin{matrix}  b^- \\ a^+ \end{matrix}\right),\;\;\;\;x\rightarrow\pm\infty,
\label{eq19}
\end{equation}
where $\mathbf{S}$ is the scattering matrix or the $S$-matrix of the propagative waves. The complex eigenvalues of the S-matrix are $\psi_{1,2}~=~T_t~\pm~R_t$. An eigenvalue of the $S$-matrix equal to zero implies that the incident wave is completely absorbed $(a^- = b^+ = 0)$. This happens when $T_t = \pm R_t$ and the incident waves $[a^+$ , $b^-]$ correspond to the relevant eigenvector. When the eigenvalues are evaluated in the complex frequency plane\cite{romero2016use}, poles and zeros can be identified. The pole frequencies correspond to the resonances of the resonator (zeros of the denominator of the eigenvalues) while the zero frequencies (zeros of the numerator of the eigenvalues) correspond to the perfect absorption configuration. In the case of a reflection problem, the eigenvalues are reduced to the reflection coefficient.

Since the systems analyzed in this work are invariant under time-reversal symmetry, the scattering matrix, as defined in Eq.(\ref{eq19}), presents unitary property\cite{norris1998reflection} in the lossless case (i.e., without dissipative losses):
\begin{equation}
\mathbf{S}^*\mathbf{S}=\mathbf{I}.
\label{eq20}
\end{equation}

The complex frequencies of the eigenvalue poles of the propagative S-matrix are complex conjugates of its zeros. Poles and zeros appear therefore symmetric with respect to the real frequency axis in the lossless case. 

\subsection{Reflection problem}

\subsubsection{Lossless case.}

\begin{figure}
\includegraphics[width=160mm]{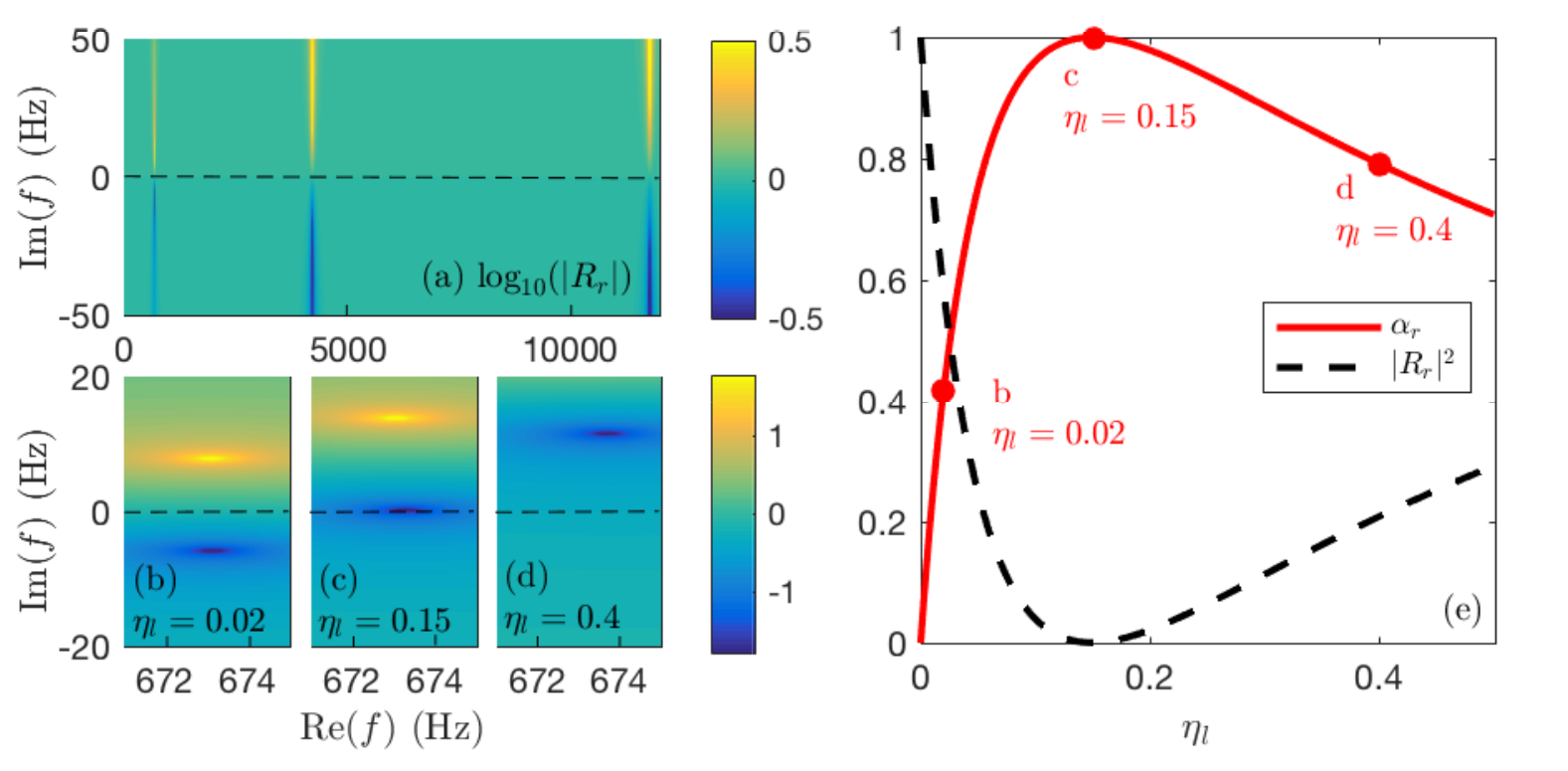}
\caption{Analysis of the scattering in the reflection problem. (a) Representation of $\log_{10}(|R_r|)$ in the complex frequency plane for the lossless case. (b)-(d) $\log_{10}(|R_r|)$ in the complex frequency plane in the lossy case for configurations with $\eta_l=0.02$, $0.15$ and $0.4$ respectively. The case when the critical coupling condition is fulfilled ($\eta_l=0.15$) is represented in (c). (e) Trade-off of the absorption at the first resonance frequency of the resonator as the inherent loss $\eta_l$ is increased in the system. The points along the absorption curve represent the values of the absorption for the configurations represented in figures~(b)-(d). Red continuous (Black dashed) line represents the absorption (reflection) coefficient as a function of $eta_l$ at 673 Hz, corresponding to the first resonance frequency of the termination.}
\label{fig:fig2}
\end{figure}

In the reflection problem, where no wave is transmitted, the reflection coefficient $R_r$ represents the scattering of the system. Thus, $R_r$ corresponds directly to both the $S$-matrix and its associated eigenvalue ($\psi=R_r$). Its zeros correspond to the cases in which the incident wave is totally absorbed. In the lossless case, $|R_r|=1$ for any purely real frequency and the pole-zero pairs appear at complex conjugate frequencies. Figure \ref{fig:fig2}a depicts $\log_{10}(|R_r|)$ in the complex frequency plane. The main beam, the resonator beam and the coating layer have the geometric and material parameters given in table \ref{table1}. Note that the Young moduli are purely real in the lossless case ($\eta=\eta_2=\eta_l=0$). As shown in section \ref{Property_S_matrix}, the poles and zeros appear in pairs and are symmetric with respect to the real frequency axis. Moreover the value of $|R_r|$ along the real frequency axis is equal to 1. It is also worth noting that the imaginary part of the pole in the lossless case represents the amount of energy leakage by the resonator through the main beam \cite{romero2016use}. With the time dependence convention used in this work, the wave amplitude at the resonance frequency decays as $e^{-\textrm{Im}(\omega_{pole})t}$. Thus the decay time $\tau_{leak}$ can be related with the quality factor due to the leakage as $Q_{leak}=\displaystyle{\frac{\textrm{Re}(\omega_{pole})\tau_{leak}}{2}=\frac{\textrm{Re}(\omega_{pole})}{2\textrm{Im}(\omega_{pole})}}$, where the leakage rate can be defined as $\Gamma_{leak}=1/\tau_{leak}=\textrm{Im}(\omega_{pole})$. The imaginary part of the poles and zeros increases when the real part of the frequency increases, meaning that more energy leaks out through the resonator at high frequencies.

\subsubsection{Lossy case.}

For the sake of clarity, this section only focuses on the first pole-zero pair of the system previously described. The discussion can nevertheless be extended to any pole-zero pair of the system in the complex frequency plane. Losses are now introduced into the system by adding an imaginary part to the Young modulus of the damping material such that it can be written as $E_{l}(1 + i\eta_l)$.

As a consequence, the symmetry between the poles and zeros with respect to the real frequency axis is broken, since the property of Eq.(\ref{eq20}) is no more satisfied in the lossy case. Figures \ref{fig:fig2}b-\ref{fig:fig2}d depict $\log_{10}(|R_r|)$ in the complex frequency plane around the first resonance frequency for three different increasing values of $\eta_l$. Figure \ref{fig:fig2}b represents the case for which the losses are small ($\eta_l=0.02$). In this case, the pole-zero pair is quasi-symmetric with respect to the real frequency axis. As the losses in the damping layer increase ($\eta_l=0.15$ in figure~\ref{fig:fig2}c and $\eta_l=0.4$ in figure~\ref{fig:fig2}d), the zero moves to the real frequency axis. In particular, the zero of the reflection coefficient is exactly located on the real frequency axis in figure~\ref{fig:fig2}c. In this situation, the amount of inherent losses in the resonator equals the amount of energy leakage. This situation is known as the critical coupling condition \cite{yariv2000universal} and implies the impedance matching, leading to a perfect absorption.

The value of the absorption coefficient of the first resonant peak as a function of $\eta_l$ is depicted in figure~\ref{fig:fig2}e. The position of the zero in the complex frequency plane is directly related to the value of the flexural wave absorption. When the zero approaches the real frequency axis, the value of the absorption is close to one, being equal to 1 when the zero is exactly located in the real frequency axis. It should be noted that the perfect absorption cannot occur once the zero has crossed the real frequency axis. This property might appear counterintuitive since it means that adding a large amount of losses in the system might lead to a deterioration of the absorbing properties of the structure.

\subsubsection{Design of perfect absorbers for flexural wave in the reflection problem.}
\label{design_reflection}

A theoretical design for the perfect absorption of flexural waves is shown in this section based on the configuration represented in the figure~\ref{fig:fig1}b and the parameters given in table \ref{table1}. Considering that there is no inherent losses in the main beam and the resonator beam ($\eta=\eta_2=0$), the loss factor of the coating layer has to be $\eta_l=0.15$ to obtain a perfect absorption at the first resonance frequency of the system.

\begin{figure}
\centering
\includegraphics[width=160mm]{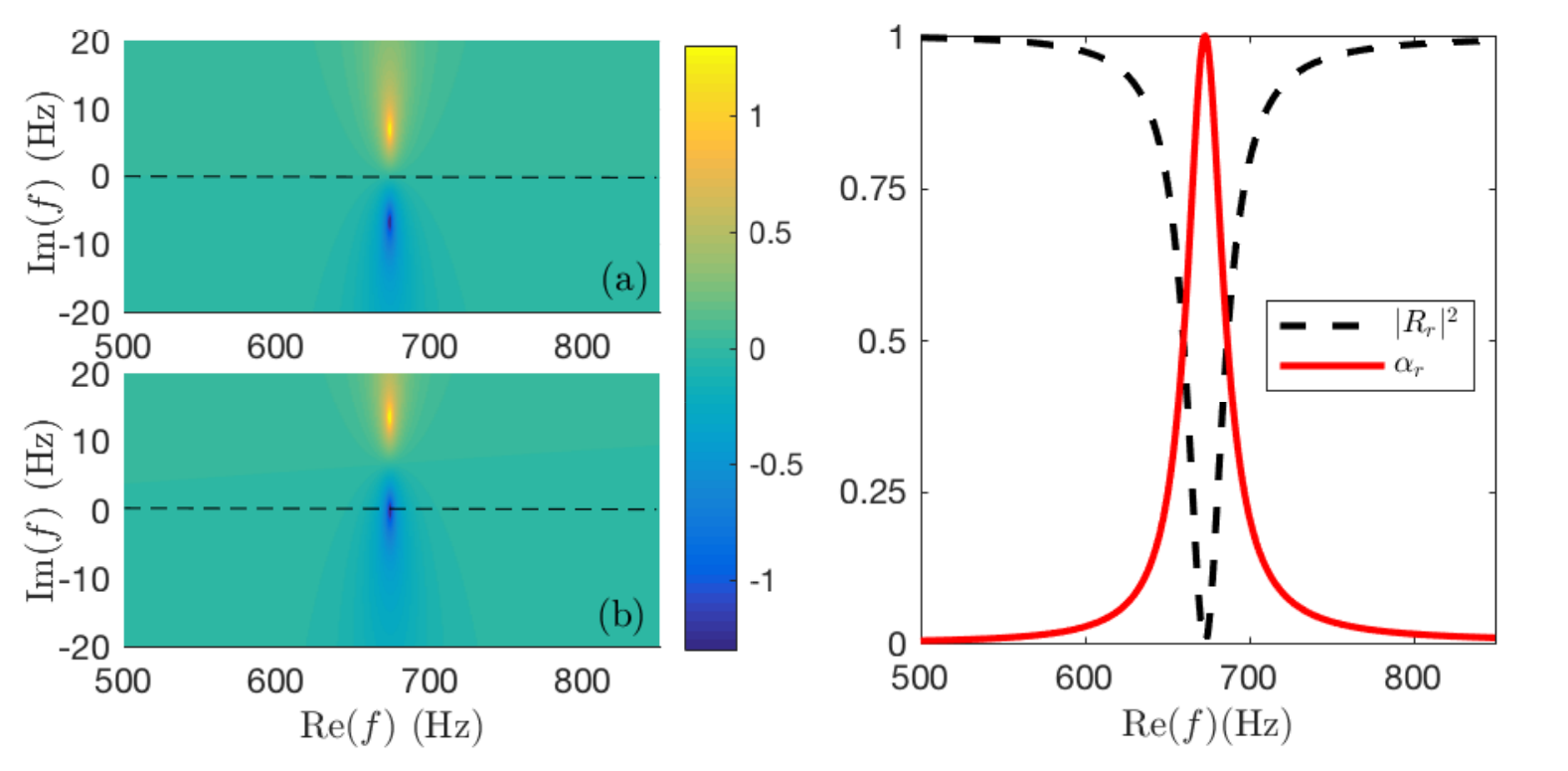}
\caption{Representation of the perfect absorption for the reflection problem. (a), (b) show the representation of the $\log_{10}(|R_r|)$ for the lossless and lossy configurations respectively. (c) Red continuous and black dashed lines show the analytical absorption and reflection coefficients for the critical coupled configuration respectively.}
\label{fig:fig3}
\end{figure}

Figures~\ref{fig:fig3}a-\ref{fig:fig3}b depict $\log_{10}(|R_r|)$ for the lossless and lossy configurations in the complex frequency plane respectively. Figure~\ref{fig:fig3}b shows particularly the first pole-zero pair of the system in the perfect absorption configuration where the critical coupling condition is fulfilled, showing the zero exactly located on the real frequency axis. Figure~\ref{fig:fig3}c shows the corresponding absorption (red continuous line) and reflection (black dashed line) coefficients according to real frequencies for the critical coupled configuration. These coefficients are calculated with the analytical model described in previous sections. The incident wave is totally absorbed at the first resonance frequency of the composite beam.

\subsection{Transmission problem}
For the transmission problem, the $S$-matrix is defined in Eq.~(\ref{eq19}) and has two eigenvalues $\psi_{1,2}$. The scatterer being mirror symmetric, the problem can be reduced to two uncoupled sub-problems: a symmetric problem where $\psi_s=T_t + R_t$ and an anti-symmetric, where $\psi_a=T_t - R_t$. 

$\psi_s$ corresponds to the reflection coefficient of the symmetric problem while $\psi_a$ corresponds to the reflection coefficient of the anti-symmetric problem. The absorption coefficient can also be expressed as $\alpha =(\alpha_s +\alpha_a)/2$ where $\alpha_s=1-\mid\psi_s\mid^2$ and $\alpha_a=1-\mid\psi_a\mid^2$. Similarly to the reflection problem, poles and zeros of $\psi_s$ and $\psi_a$ can be identified in the complex frequency plane. The following sections focuses on the first resonant mode of the beam resonator, the displacement distribution of which is symmetric. The interpretation of the results can nevertheless be applied to the higher order modes with anti-symmetric distributions of the displacement field. It is worth noting that the displacement distribution of the resonant modes changes from symmetric to anti-symmetric as the mode increases due to the geometry of the resonators \cite{Jimenez2017}.

\subsubsection{Lossless case.}
\begin{figure}
\includegraphics[width=160mm]{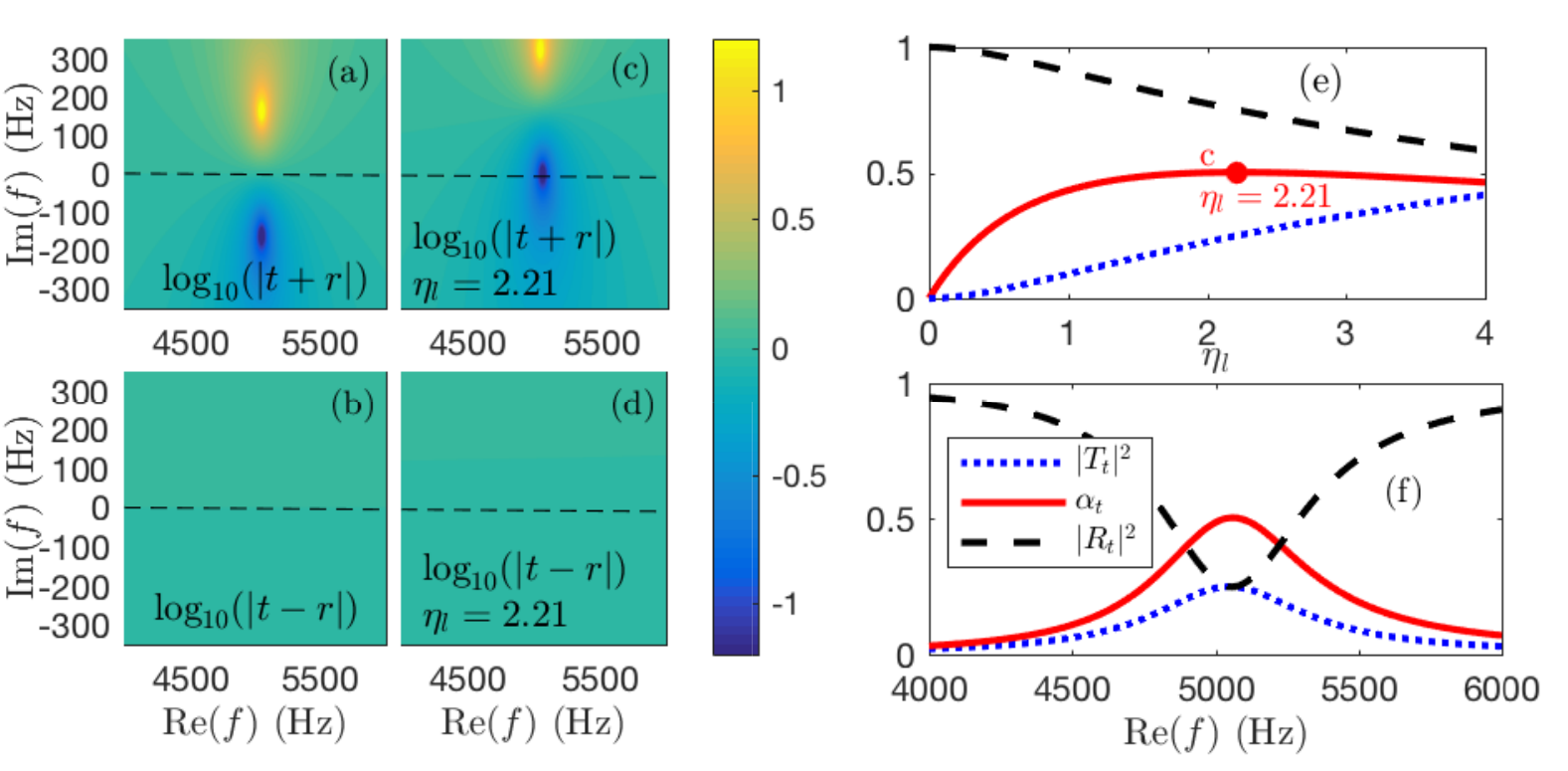}
\caption{Representation of the eigenvalues of the $S$-matrix for a transmission problem in the lossless and lossy case. (a) and (b) Lossless case for $\log_{10}(|\psi_s|)$ and $\log_{10}(|\psi_a|)$ in the complex frequency plane. (c) and (d) Lossy case for $\log_{10}(|\psi_s|)$ and $\log_{10}(|\psi_a|)$ in the complex frequency plane. (e) Trade-off of the transmission (blue dotted line), reflection (black dashed line) and absorption (red continuous line) for the maximum absorption of the first mode as the loss factor of the coating layer increases. (f) Red continuous, black dashed and blue dotted lines represent the absorption, reflection and transmission coefficients respectively for the half critically coupled configuration.}
\label{fig:fig4}
\end{figure}

Figures~\ref{fig:fig4}a and \ref{fig:fig4}b show the variation of $\log_{10}(|\psi_s|)$ and $\log_{10}(|\psi_a|)$ evaluated respectively in the complex frequency plane in the lossless case for the first resonant mode. The main beam, the resonator beam and the coating layer of the studied system have still the material and geometric parameters of table \ref{table1}, where $\eta=\eta_2=eta_l=0$ in the lossless case. The symmetric and anti-symmetric problems exhibit pole-zero pairs similarly to the reflection problem in the lossless case. These pairs are also symmetrically positioned with respect to the real frequency axis. The absence of dissipation is shown along the real frequency axis where $|T_t\pm R_t|=1$ for any real frequency. This section focuses only on the first resonant mode which has a symmetric distribution of the displacement field. Therefore, only the symmetric problem presents a pole-zero pair at the corresponding resonance frequency, while the anti-symmetric one does not. 

\subsubsection{Lossy case.}
\label{sec:lossy}
Unlike the reflection problem, the condition for perfect absorption is stronger in the transmission one and needs to place the zeros of both $\psi_s$ and $\psi_a$ at the same frequency in the real frequency axis. Once this condition is fulfilled, $a^+$ and $b^-$ correspond to the relevant eigenvector and the system satisfies the Coherent Perfect Absorption (CPA) condition \cite{chong2010coherent, Wei14, merkel2015control}.

Losses are introduced in the system in the same way as for the reflection problem, i.e., by increasing the loss factor $\eta_l$ of the material of the damping layer. Once the losses are introduced, the position of the pole-zero pair of the symmetric eigenvalue in the complex frequency plane shifts towards the upper half space while the anti-symmetric problem remains unchanged without pole-zero pairs, as shown in Figures~\ref{fig:fig4}c and \ref{fig:fig4}d. Therefore, only the zero of the symmetric problem can be placed on the real frequency axis, i.e., only half of the problem can be critically coupled. Figure~\ref{fig:fig4}e shows the dependence of the reflection, transmission and absorption coefficient on the inherent losses in the resonator for the first mode. The maximum absorption obtained is 0.5 since only the symmetric problem is critically coupled ($\alpha =(\alpha_s +\alpha_a)/2\simeq(1+0)/2=1/2$).

\subsubsection{Design of maximal absorbers for flexural wave in the transmission problem.}

Based on the results discussed previously, a configuration with maximal absorption for flexural waves in the transmission problem is designed with the parameters given in table \ref{table1}. As for the reflection problem, no inherent losses are considered in the main beam and the resonator beam ($\eta=\eta_2=0$). The loss factor of the coating layer is $\eta_l=2.21$. The reflection, transmission and absorption for this configuration is analyzed in figure~\ref{fig:fig4}(f), showing that the maximum absorption is 0.5 at the resonance frequency of the beam. This result is in accordance with the ones previously obtained \cite{El-Khatib05,Wei14, merkel2015control}, even if the resonator is not a point translational impedance. The absorption is limited to 0.5 since only one kind of geometry of resonant mode can be excited. The problem is therefore half critically coupled. To obtain a higher absorption, other strategies based on breaking the symmetry of the resonator \cite{jimenez2017rainbow} or on the use of degenerate resonators are needed \cite{yang2015}. In these cases, both eigenvalues present poles and zeros located at the same real frequencies. It would then be possible to fully critically couple the problem and so obtain a perfect absorption (i.e. $\alpha=1$) at the appropriate frequency.

\section{Experimental results}
\label{sec:exp1}
This section presents the experimental results of the reflection coefficient \cite{gautier2006reflection,denis2015measurement} for an aluminum beam system with the configuration described in section \ref{design_reflection}. 

\subsection{Experimental set-up}
\label{sec:exp}

\begin{figure}
\includegraphics[width=16cm]{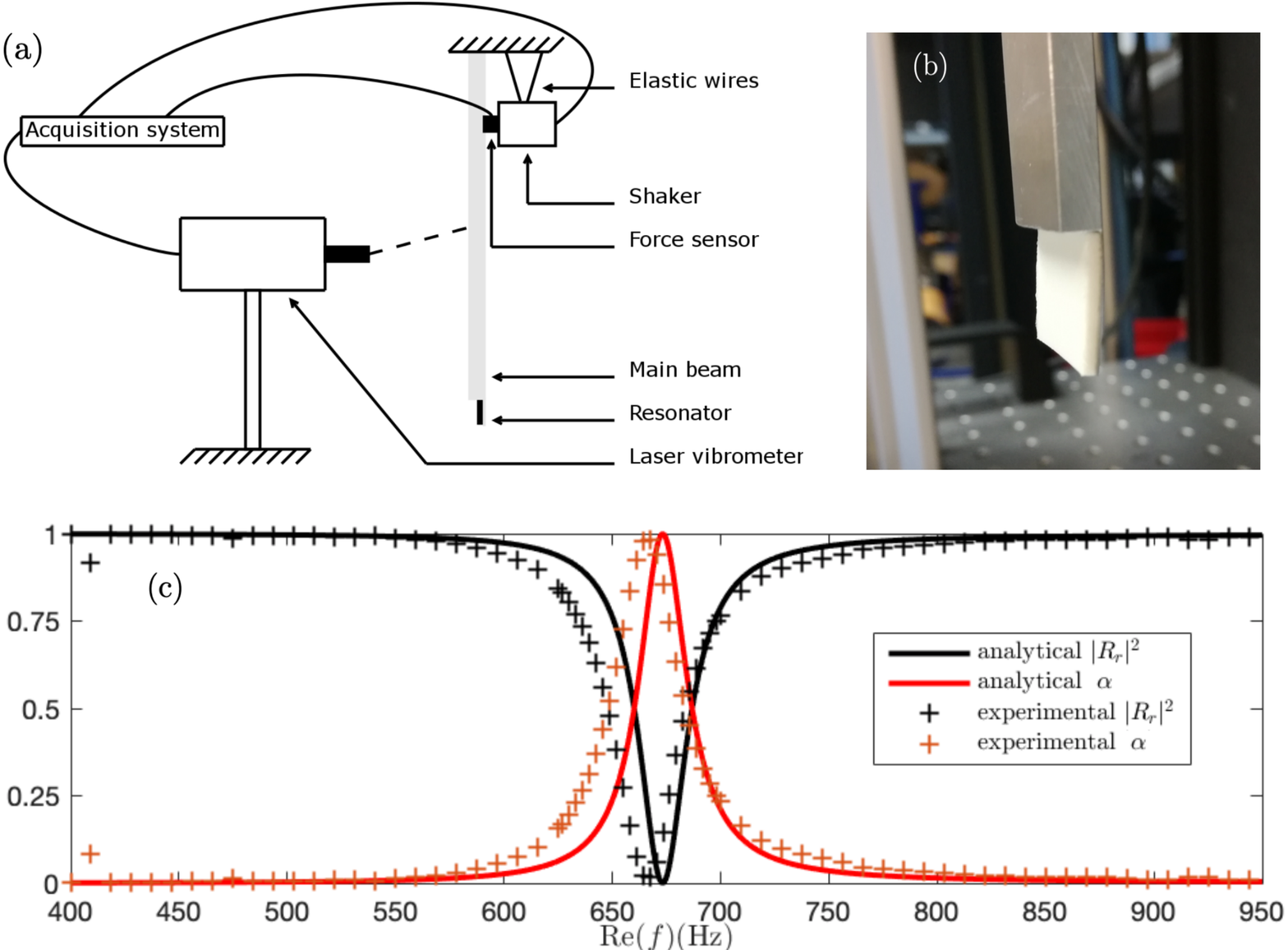}
\caption{(a) Diagram of the experimental set-up. (b) Photograph of the resonator. (c) Black crosses and red open circles show respectively $|R_r|^2$ and $\alpha_r$ for the critical coupled configuration measured with the experimental set-up. Black dashed and red continuous lines show $|R_r|^2$ and $\alpha_r$ calculated with the analytical model.}
\label{fig:fig5}
\end{figure}

The beam is held vertically in order to avoid static deformation due to gravity. The extremity at which the reflection coefficient is estimated is oriented towards the ground (see figure~\ref{fig:fig5}a). The used coating layer have been experimentally characterized showing an $\eta_l=0.15$, which is the value for which perfect absorption can be observed.  A photograph of the resonator with the coating layer is shown in figure~\ref{fig:fig5}b. The measurements are performed along the beam at 21 points equidistant of 5~mm and located on its neutral axis in order to avoid the torsional component. The measurement points are also located sufficiently far from the source and the extremity of the beam to consider far-field assumption and neglect the contribution of evanescent waves. In this case, far-field assumption is fulfilled at a distance $l_f$ from both the source and the resonator for which the evanescent wave loses 90 $\%$ of its initial magnitude. The low frequency limit of the measurements is then estimated using $e^{kl_f}=0.1$. The shaker excites the beam with a sweep sine. The displacement field versus frequency is obtained from the measurements of the vibrometer at each measure point.

\subsection{Experimental estimation of the reflection coefficient}

Consider the flexural displacement $W(x_i,\omega)$ measured at the point $x_i (i \in [0 ,20])$ for a given angular frequency $\omega$ as
\begin{equation}
W(x_i,\omega)=A(\omega) e^{-ik x_i}+ B(\omega)  e^{ik x_i}.
\end{equation}
The set of $W(x_i,\omega)$ for each measurement point can be written in a matrix format \cite{denis2015measurement} such as
\begin{equation}
\left(\begin{matrix}
W(x_0,\omega)\\
W(x_1,\omega)\\
W(x_2,\omega)\\
\vdots \\
W(x_{20},\omega)
\end{matrix}
\right) = \left(\begin{matrix}
e^{(-ik x_0)} & e^{(ik x_0)}\\
e^{(-ik x_1)} & e^{(ik x_1)}\\
e^{(-ik x_2)} & e^{(ik x_2)}\\
\vdots & \vdots \\
e^{(-ik x_{20})} & e^{(ik x_{20})}\\
\end{matrix}
\right)
\left(\begin{matrix}
A(\omega)\\
B(\omega)\\
\end{matrix}
\right),
\label{eq23}
\end{equation}
The amplitudes $A(\omega)$ and $B(\omega)$ can then be derived from Eq.~(\ref{eq23}) which forms an overdetermined system. From these amplitudes, the reflection coefficient of the propagative waves can be deduced for any $\omega$ as:
\begin{equation}
R_r(\omega)= \frac{A(\omega)}{B(\omega)}.
\end{equation}

\subsection{\label{experimental_evidence} Experimental evidence of perfect absorption for flexural waves}

Experimental results obtained with the experimental set-up are depicted in figure~\ref{fig:fig5}c. A drop of reflection is noticed at the first resonant frequency of the termination with a minimum value of $|R_r|^2=0.02$ at $667$ Hz for the experiment and $|R_r|^2=0$ at $673$ Hz for the analytical result. The gap between the analytical and experimental resonant frequency is $0.9\%$. This frequency shift between the model and the experiment is mainly due to the geometric uncertainty in the resonator thickness, induced by the machining process. This geometrical uncertainty induces also an estimation uncertainty of the energy leakage of the resonator. The absorption is then experimentally limited to $\alpha_r= 0.98$. Evidence of perfect absorption for flexural waves by means of critical coupling is shown experimentally here.

Three experimental scans of the whole beam at 500 Hz, 670 Hz and 800 Hz have been measured \footnote{See supplementary material: videos 500Hz.avi, 670Hz.avi and 800Hz.avi}. At 500 Hz or 800 Hz the reflection coefficient is very close to one. The standing waves in the main beam are visible at these frequencies. At 670 Hz, the termination absorbs totally the incident waves. There is therefore no standing waves and the waves are propagating in the main beam. 

\section{Conclusions}
\label{sec:con}

Absorption of propagative flexural waves by means of simple beam structures is analyzed in this work. The main mechanisms are interpreted in terms of both the critical coupling conditions and the symmetries of the resonances for both the reflection and the transmission problems. The positions of the zeros of the eigenvalues of the scattering matrix in the complex frequency plane give informations on the possibility to obtain the perfect absorption. The perfect absorption condition is fulfilled when these zeros are placed on the real frequency axis, meaning that the inherent losses are completely compensating the energy leakage of the system. In the reflection problem, the physical conditions of the problem lead to perfect absorption at low frequencies. In this case a single symmetry for the resonance is excited and perfect absorption can be obtained when the inherent losses of the system balance the energy leakage of the system. In the transmission problem, the requirement to obtain perfect absorption is stronger than for the case in reflection as two kinds of symmetries of the resonances are required to be critically coupled simultaneously. In the case presented in this work, or in the general case of point translational impedances, dealing only with one type of symmetry for the resonant modes \cite{El-Khatib05} limits the absorption to 0.5. Therefore for the perfect absorption in the transmission case, two strategies are needed: ($i$) breaking the symmetry of the resonator in order to treat the full problem with a single type of symmetry of the resonance mode \cite{jimenez2017rainbow}; ($ii$) using degenerate resonators with two types of symmetries at the same frequency being critically coupled \cite{yang2015}. The resonator used in this study has been chosen as an integral part of the main beam for experimental set-up  reasons. However, the presented approach can be applied to any class of 1D resonant-system provided that the resonators are local, open and lossy ones. These properties of the resonator are the essential points to achieve the perfect absorption at low frequency by solving the following problems: increasing the density of states at low frequencies and matching the impedance with the background medium.

\section*{Acknowledgments}	
The authors thank Mathieu S\'ecail-Geraud and Julien Nicolas for the development of the experimental set-up. The work has been founded by the RFI Le Mans Acoustic (R\'egion Pays de la Loire) within the framework of the Metaplaque project. This article is based upon work from COST action DENORMS CA 15125 , supported by COST ( European Cooperation in Science and Technology). The work was partly supported by the Spanish Ministry of Economy and Innovation (MINECO) and European Union FEDER through project FIS2015-65998-C2-2 and by project AICO/2016/060 by Conseller\'ia de Educaci\'on, Investigaci\'on, Cultura y Deporte de la Generalitat Valenciana.	

\section*{References}

\end{document}